\documentclass[reprint, superscriptaddress, amsmath, amssymb, aps, prl]{revtex4-2}

\usepackage[utf8]{inputenc}
\usepackage[english]{babel}
\usepackage[T1]{fontenc}
\usepackage{amsmath,amssymb}
\usepackage{times}
\usepackage{siunitx} 
\usepackage[normalem]{ulem} 
\usepackage{import}
\usepackage{graphicx} 
\usepackage{hyperref}[breaklinks=true]

\usepackage{lipsum}
\usepackage{xcolor} 

\newcommand{\TUWien}{Vienna Center for Quantum Science and Technology, Atominstitut, TU Wien, Stadionallee 2, 1020 Vienna, Austria}

\begin{document}

\title{Experimental Observation of 
Curved Light-Cones in a Quantum Field Simulator}

\author{Mohammadamin~Tajik}
\email{amintajik.physics@gmail.com}
\affiliation{\TUWien}%
\author{Marek~Gluza}
\affiliation{School of Physical and Mathematical Sciences, Nanyang Technological University, 639673 Singapore, Republic of Singapore}
\author{Nicolas~Sebe}
\affiliation{\TUWien}%
\affiliation{Dahlem Center for Complex Quantum Systems, Freie Universit{\"a}t Berlin, Berlin, Germany}
\affiliation{École Polytechnique, Route de Scalay, 91120 Palaiseau, France}
\author{Philipp~Sch{\"u}ttelkopf}
\author{Federica~Cataldini}
\affiliation{\TUWien}
\author{Jo{\~a}o~Sabino}
\affiliation{\TUWien}
\affiliation{Instituto Superior T\'{e}cnico, Universidade de Lisboa, Av. Rovisco Pais 1, 1049-001, Lisbon, Portugal}
\author{Frederik~M{\o}ller}
\author{Si-Cong~Ji}
\affiliation{\TUWien}
\author{Sebastian~Erne}
\affiliation{\TUWien}%
\author{Giacomo~Guarnieri}
\affiliation{Dahlem Center for Complex Quantum Systems, Freie Universit{\"a}t Berlin, Berlin, Germany}
\author{Spyros~Sotiriadis}
\affiliation{Dahlem Center for Complex Quantum Systems, Freie Universit{\"a}t Berlin, Berlin, Germany}
\author{Jens~Eisert}
\affiliation{Dahlem Center for Complex Quantum Systems, Freie Universit{\"a}t Berlin, Berlin, Germany}
\affiliation{Helmholtz-Zentrum Berlin f{\"u}r Materialien und Energie, 14109 Berlin, Germany}
\author{J{\"o}rg~Schmiedmayer}
\email{schmiedmayer@atomchip.org}
\affiliation{\TUWien}

\begin{abstract}
We investigate signal propagation in a quantum field simulator of the Klein-Gordon model realized by two strongly coupled parallel one-dimensional quasi-condensates. By measuring local phononic fields after a quench, we observe the propagation of correlations along sharp light-cone fronts. If the local atomic density is inhomogeneous, these propagation fronts are curved. For sharp edges, the propagation fronts are reflected at the system’s boundaries. By extracting the space-dependent variation of the front velocity from the data, we find agreement with theoretical predictions based on curved geodesics of an inhomogeneous metric. This work extends the range of quantum simulations of non-equilibrium field dynamics in general spacetime metrics.
\end{abstract}

\maketitle


\textit{Introduction}.---Light-cones embody one of the most fundamental principles in physics: Causality. When constructing models describing fundamental interactions in nature, one of the basic requirements is the existence of light-cones. Indeed, it has been understood that they appear as a result of the  relativistic invariance of quantum fields~\cite{peskin1995introduction}.
Interestingly, there are several systems whose effective dynamics are relativistically invariant, and effective light-cones also play a role. Recent experiments have revealed that effective light-cones do emerge in cold atomic gases~\cite{cheneau2012light,lightcone0}.
In order to directly observe these light-cones, several experimental challenges had to be overcome, including resolving the system at fine length-scales and measuring relevant observables that would be able to reveal them.
Tackling such issues is part of a larger research endeavor on devising quantum simulators~\cite{CiracZollerSimulation,1408.5148}. 
For example, manipulation of one-dimensional tunnel-coupled gases allows the simulation of prototypical field theories \cite{Schweigler_SG,zachePhysRevX.10.011020,Recurrence,LongGaussification} that are of foundational importance but also, e.g., capture charge transport in nano-wires~\cite{Giamarchi2004}. 
Here, our goal is to use this quantum simulator to explore experimentally its potential to simulate dynamics in inhomogeneous or curved metrics.
Similar objectives have been the focus of analogue gravity systems \cite{barcelo2011analogue} which recently have been very successful in simulating black hole \cite{munoz2019observation,kolobov2021observation} or cosmological \cite{Jaskula2012_Casimir,eckel2018rapidly,Oberthaler_2022} processes using cold-atom systems. 

In this work, we investigate the correlation propagation in an inhomogeneous one-dimensional quantum gas. We show that correlation fronts follow geodesics of the analogue acoustic metric and find the spatial dependence of the propagation velocity in agreement with the theoretical modeling. We observe ballistic propagation of correlation fronts and discuss the detailed shape, reflections at the system's boundaries, and periodic recurrences of these correlation fronts.

\emph{Quantum field simulation}.---We use two tunneling-coupled one-dimensional superfluids~\footnote{See Supplemental Material for further details.} to simulate the inhomogeneous Gaussian field theory in $1+1$ spacetime dimensions whose action for a bosonic field $\phi$ can be written as
\begin{equation}
    \mathcal{S}[\phi]
    \! \sim \! \!
    \int \! \! \mathrm{d}z\mathrm{d}t 
    \sqrt{-g} K(z) 
    \left[g^{\mu\nu} \! \left(\partial_\mu \phi\right) \! \left(\partial_\nu \phi  \right) 
    + \tfrac{1}{2} M^2 {\phi}^2 \right]\, ,
     \label{eq:action}
\end{equation}
where $M$ is the mass and $g=\det (g_{\mu \nu})$. The spacetime interval $\mathrm{d}s$ of the metric tensor $g_{\mu\nu}$ is given by 
\begin{equation}\label{eq:metric}
    \mathrm{d}s^2 = 
    g_{\mu\nu}\mathrm{d}x^\mu\mathrm{d}x^\nu 
    = - v(z)^2 \mathrm{d}t^2 + \mathrm{d}z^2\,. 
\end{equation}
Here, in accordance with the presented experiment, we neglected an explicit time-dependence of the parameters $v$ and $K$.
Light-like trajectories in this metric deviate from straight lines according to the function $v(z)$, i.e., the local propagation speed of (massless) fluctuations.

Note that, due to the conformal invariance of the Laplacian in $1+1$ dimensions, the scale factor $K(z)$ cannot be absorbed into the metric by a conformal transformation and hence has to be included in the action for generality. Nevertheless, we find the propagation of correlation fronts to be dominated by the induced metric, i.e.,~the massless Klein-Gordon equation, and hence neglect the spatial dependence of $K(z)$ for simplicity (see
Ref.~\cite{Note1} for details).

\begin{figure*}[t]
    \centering
   \includegraphics[width=\linewidth]{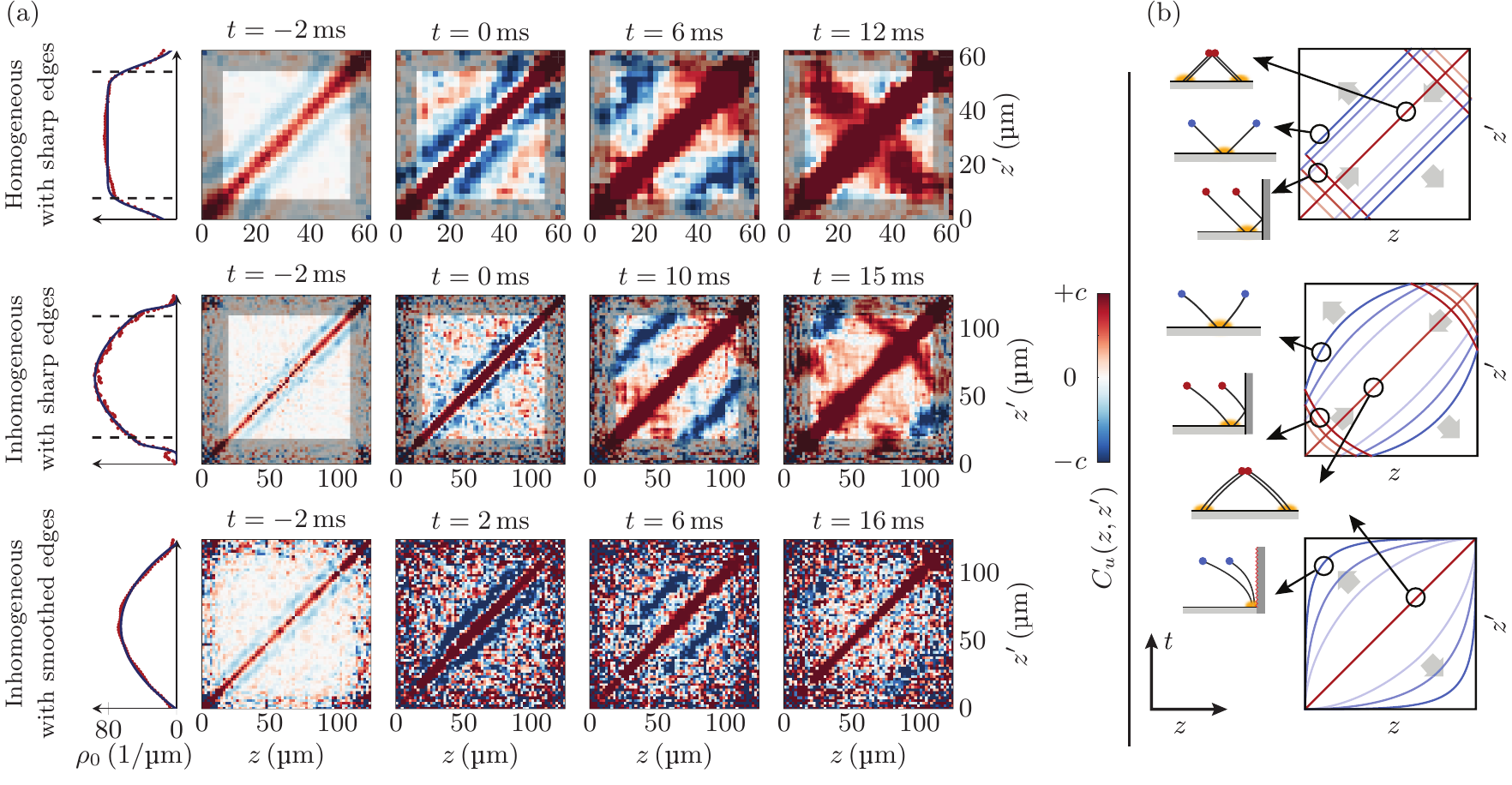}
    \caption{ Propagation of the flat and curved fronts of the two-point correlation function of the velocity field $\hat u$. {\bf (a)} Measurement results of $C_u(z,z^\prime)$ at selected times $t$ for three experimental settings with different background density profiles as explained on the left. For each case, the measured $\rho_0(z^\prime)$ is presented by red dots and the blue line is a fit. {\bf (b)} Intuitive explanation of the correlation propagation fronts in different cases. 
    At $t=0$, the points marked with the yellow dots are correlated. After the quench, evolution of correlations between these two points is traced by the light-cone trajectories of the left- and right-moving chiral fields, as depicted in the spacetime sketches.  
      The presence of narrow width fronts, as well as their shape and sign of correlations are fully explained by taking into account the effects of reflection at the boundaries and effectively curved metric. 
      For further details, see Ref.~\cite{Note1}. 
      \label{fig:main}}
\end{figure*}

In our experiment, the bosonic field $\phi$ corresponds to the relative phase between the two superfluids and the space-dependent speed of sound is related to the local averaged density $\rho_0(z)$ of each superfluid, through
\begin{align}
    v(z)=\sqrt{{g_\mathrm{1D}\rho_0(z)}/{m}}\, ,
    \label{eqsoundvelocitymain}
\end{align}
where $m$ is the mass of an atom, and $g_\mathrm{1D}$ the effective  inter-atomic interaction strength. Thus, local tuning of the density by changing the trapping potential~\cite{DMD}, allows for modification of the key physical parameter in the simulated metric Eq.~\eqref{eq:metric}.

To initiate the light-cone propagation of correlations, we perform a quench by rapidly changing the mass parameter $M$ from an initially large value to zero. Such a mass quench is a paradigmatic protocol for inducing non-equilibrium dynamics of a quantum field~\cite{CalabreseCardy06}. In the experiment, the mass $M$ is related to the single atom tunneling rate between the two atomic clouds and is quenched to zero by separating the superfluids and letting them evolve independently~\cite{Schumm05,Schweigler_SG}.

To directly observe the light-cone dynamics, we need to measure the correlations of a local observable, like the particle density or current, which are the fundamental fields in the effective hydrodynamic description. 
The particle current at position $z$ is related to the spatial derivative of the phase field, $\hat{j}(z)= \rho_0(z) \hat{u}(z)$ with the fluid velocity field
\begin{align}
    \hat{u}(z)=(\hbar/m)\partial_{z}\hat{\phi}(z)\ .
    \label{eqvelocityfield}
\end{align}
We measure the velocity field by extracting the spatially resolved relative phase through matterwave interferometry and show that after the quench, correlations of the velocity field exhibit light-cone fronts allowing to explore aspects of the quantum-simulated spacetime geometry.

\begin{figure}[t]
    \centering
    \includegraphics[width=0.9\linewidth]{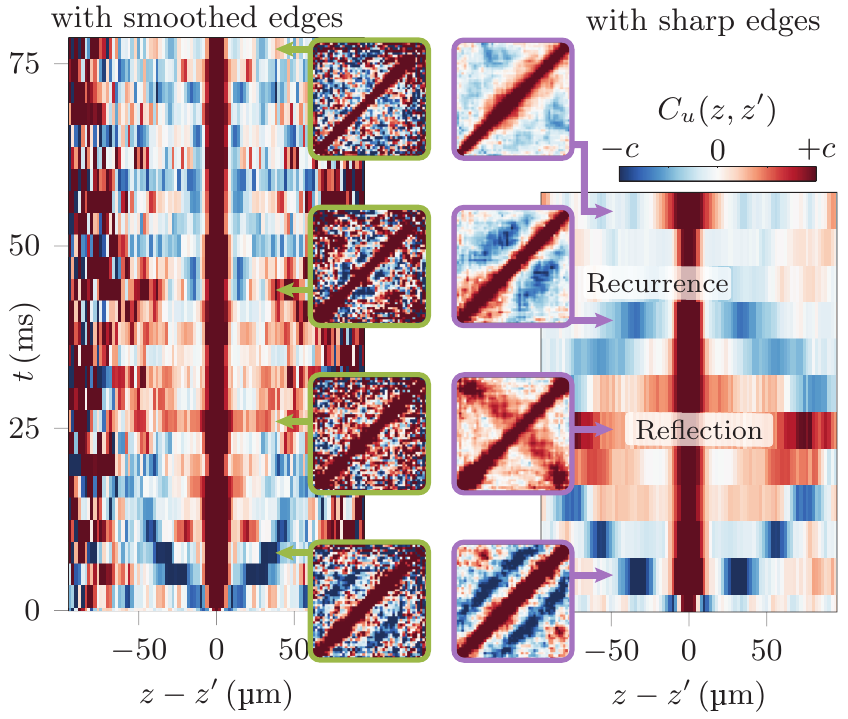}
    \caption{Observation of light-cone propagation in the two-point correlation function of the hydrodynamic velocity field, $C_u(z,z^\prime=-z)$, in experimental settings with (right) and without (left) sharp edges. In the two main plots, the anti-diagonal correlations are plotted over time. For the density with sharp edges, reflection from boundaries and the recurrence is clearly observed, which is missing in the other case. For selected time steps, the full two-dimensional correlation function is plotted similar to Fig.~\ref{fig:main}.}
    \label{fig:reflection}
\end{figure}

\emph{Experimental results}.---In Fig.~\ref{fig:main}(a)~and~\ref{fig:reflection}  we show the dynamics of the two-point correlation functions of the velocity field, $C_u(z,z^\prime) = \langle \hat{u}(z) \hat{u}(z^\prime) \rangle$, at different times $t$ for three different experimental settings: A homogeneous density with sharp edges, an inhomogeneous density also with sharp edges, and an inhomogeneous density with smoothed edges, presented in Fig.~\ref{fig:main} from top to bottom respectively. All density profiles $\rho_0(z)$ are shown in Fig.~\ref{fig:main}(a). 

In accordance with Eq.~\eqref{eq:metric}, assuming a vanishing macroscopic background particle flow, we find the average current $\langle \hat u(z)\rangle = 0$ throughout the evolution. Nevertheless,  nonzero correlations  $C_u(z,z^\prime) \neq 0$ mean that current fluctuations are not independent between $z$ and $z^\prime$: If the correlation is positive, then in individual measurements, the current fluctuations at $z$ and $z^\prime$ tend to be aligned. Conversely, for negative correlation the current fluctuations point in opposite directions.
Initially, for the thermal state with large $M$, the velocity only has short range correlations, with $C_u(z,z^\prime)$ consisting of auto-correlations on the diagonal (red)  accompanied by anti-correlations (blue) parallel to the diagonal \cite{Note1}.

After the quench, we find that the propagation of correlation fronts is determined by the geodesics of the induced spacetime metric, and hence can be understood in a phenomenological quasi-particle picture \cite{PhysRevLett.96.136801, Calabrese2007,exactrelaxation}. In this picture, the dynamics is carried by pairs of initially short range correlated quasi-particles moving in opposite directions as illustrated in Fig.~\ref{fig:main}(b).
In all three experimental settings, the auto-correlations remain intact on the diagonal throughout the evolution, due to initially correlated co-moving quasi-particle pairs (see Fig.~\ref{fig:main}(b)).

The evolution of the anti-correlation fronts, on the other hand, is determined by the propagation of initially correlated counter-moving quasi-particle pairs and hence propagate away from the diagonal. As illustrated in the spacetime sketches of Fig.~\ref{fig:main}(b), this demonstrates the spreading of correlations to longer length scales due to the separation of initially correlated quasi-particle pairs. 
When the density profile is homogeneous, the anti-correlation fronts are consistent with straight lines throughout the dynamics.
For the inhomogeneous density profiles the anti-correlation fronts curve up over time which is a key qualitative effect of the quantum-simulated curved metric.
Additional effects can arise in finite size systems, due to possible reflections of fluctuations at the boundaries \cite{Essler2016}.

In cases with sharp boundaries (first two rows of Fig.~\ref{fig:main}), we observe the formation of perpendicular (anti-diagonal) fronts propagating inwards from the system boundaries. 
Unlike the fronts parallel to the diagonal, the perpendicular ones correspond to positive correlations.
The sign change is consistent with a reflection of the direction of the current fluctuation at the boundary of the system, i.e.,~the change in sign of the velocity for quasi-particles scattered at the boundary (see the reflected trajectories in Fig.~\ref{fig:main} (b)).
Note, that for reflecting boundaries the perpendicular front spans the entire system (see, e.g.,~in the homogeneous case at $t =  \SI{12}{\milli\second}$).
This reveals a new insight about the dynamics in the system: The quench dynamics have transformed the initial thermal correlations into a configuration similar to the so-called `rainbow'~\cite{ramirez2015entanglement}.

In the case with soft boundaries (third row in Fig.~\ref{fig:main}), no anti-diagonal fronts appear, signaling the absence of reflections. This is in accordance with the presence (absence) of reflections for the off-diagonal anti-correlation fronts for sharp (smoothed) edges, presented in Fig.~\ref{fig:reflection}. In the presence of sharp edges, the anti-correlation fronts change direction and return to their initial position, resulting in an approximate but clearly visible recurrence of the correlations (c.f.~\cite{Recurrence}). In contrast, for the soft boundary, we observe the slow-down of the anti-correlation fronts and the absence of reflections and recurrences at longer times.

\emph{Theoretical interpretation}.---The Gaussian field theory in an inhomogeneous metric, i.e.~Eq.~\eqref{eq:action} for $K(z) \equiv K$, allows for an exact geometric explanation of correlation front dynamics. 
The massless unitary evolution of the velocity field in homogeneous infinite space is a superposition of two local `chiral' components $\hat \chi_\pm$ evaluated at counter-propagating locations
\begin{align}
    \hat{u}(z,t)=\hat{\chi}_{+}(z+ vt)+\hat{\chi}_{-}(z- vt)\ .
\end{align} 
Thus, the time-dependent correlations can be derived by tracing the chiral components $\hat\chi_\pm$ from their origin at $t=0$. 
The initial correlation length is short, so at $t=0$ two-point correlations $\langle\hat \chi_\sigma(z)\hat\chi_{\sigma'}(z')\rangle$ with $\sigma,\sigma'=\pm$ can be significant only for nearby points $z\approx z'$ (c.f.~Fig.~\ref{fig:main}). During the dynamics this translates to the condition $\left|z-z' \right|\approx 2vt$ which corresponds to the positions of the anti-correlation fronts in the experiment. Thus, we expect the anti-correlation front to propagate at twice the sound velocity.

We discuss the conditions for the other fronts and the modification of the above calculation accounting for a finite size system with sharp edges in Ref.~\cite{Note1}.
The sign switching after reflections is consistent with an effective boundary condition of the Neumann type for the phase field, i.e., vanishing of the velocity field at the edges \cite{Patrick22_BH}. This is the right choice of boundary conditions for an atomic gas trapped in a box-like potential, as the particle current $\hat{j}$ vanishes at the edges. 

When tracing the positions in the inhomogeneous case we need to account for the space-dependent sound velocity.
To this end we replace time in the equations describing the positions of the fronts by the actual traveling time for a particle to propagate through a given space interval
\begin{equation}
    \tau(z,z')=\int_{z}^{z'}\frac{\mathrm{d}l}{v(l)}\ .
    \label{eq:s}
\end{equation}
The anti-correlation front consists of points $z, z'$ satisfying  $|\tau(z,z')|\approx 2t$ which generalizes the condition for the flat metric (see Ref.~\cite{Note1} for analogous, though more complicated, relations for the reflected fronts). 
Correlation fronts are therefore curved instead of straight lines.

\begin{figure}[t]
    \centering
    \includegraphics[width=\linewidth]{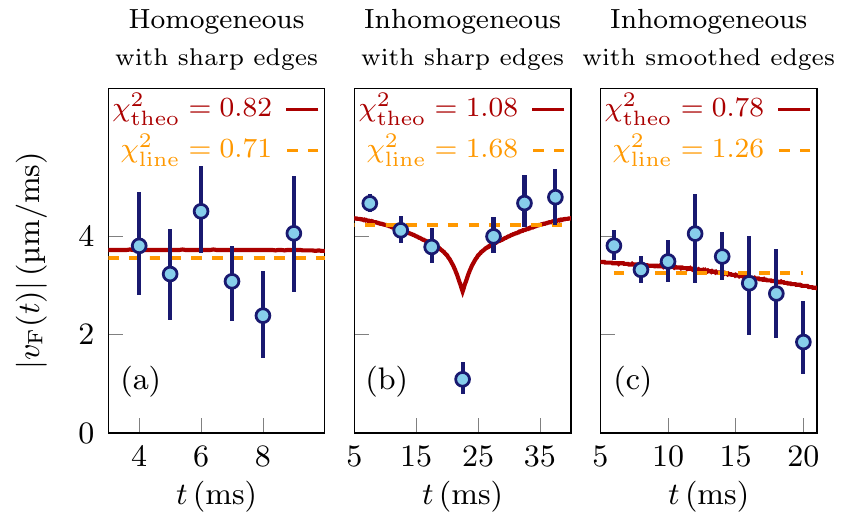}
    \caption{Estimation of the average front velocity for three settings introduced in Fig. \ref{fig:main}. The extracted velocities are shown with blue bullets. The error bars showing 68\% confidence intervals obtained via bootstrapping~\cite{efron1986bootstrap}. The red lines represent twice the speed of sound calculated from the experimental local density, $\rho_0(z)$. The orange line is a vertical line marking the average value of the blue bullets. The $\chi^2$ is the reduced chi-squared values comparing red/orange curves with the blue bullets~\cite{Note1}.}
    \label{fig:velocity}
\end{figure}

From the experimental data, we quantify the position-dependence of the effective front velocity. We estimate the front location in cuts perpendicular to the diagonal ($z^\prime= -z$) at different propagation times $t$ and compute the average front velocity $v_\mathrm{F}(t)$ via the difference quotient. In Fig.~\ref{fig:velocity}, we compare the measured $v_\mathrm{F}(t)$ (blue bullets) with the theoretical prediction (red line) and a constant velocity (orange line).
In the homogeneous case, the measured front propagation is consistent with a constant velocity (a horizontal line in Fig.~\ref{fig:velocity}(a)).
For the inhomogeneous density profiles, shown in Fig.~\ref{fig:velocity}(b)~and~(c), the effective velocities depend on position in accordance with the theoretically predicted inhomogeneous metric. 

The reflection of the correlation front is also clearly visible in Fig.~\ref{fig:velocity}(b) and we find reasonable agreement to the free Gaussian model Eq.~\eqref{eq:action}. Therein, if the velocity decreases slower than linearly towards the boundary, the traveling time diverges, so there is neither reflection nor turn. Therefore, light-like trajectories converge asymptotically to the boundary, in agreement with the geodesics of Eq.~\eqref{eq:metric}. This is in agreement with experimental observations (see Fig.~\ref{fig:reflection} and Fig.~\ref{fig:velocity}(c)). Note however, that the Luttinger liquid description is expected to break down near the edges due to vanishingly low atomic density \cite{Stringari_Bose_gas_edges} and hence the absence of reflections might be dominated by dispersive or higher-order corrections to the Gaussian model Eq.~\eqref{eq:action}.

\emph{Conclusion}.---Going beyond previous studies of tunneling-coupled one-dimensional gases~\cite{lightcone0,lightcone2,langen2015experimental,double_lightcone,Recurrence,ShortGaussification}, our results for the velocity field correlations provide a direct measurement of the underlying light-cone propagation. Controlling the local propagation speed of fluctuations $v(z)$, by shaping a stationary inhomogeneous average density, we investigated the propagation of correlation fronts in three distinct settings. In all cases, the experimentally observed light-cone propagation in our quantum field simulator was in good agreement with theoretical predictions for a bosonic Gaussian field theory in the analogue 
metric $g_{\mu \nu}$ (Eq.~\ref{eq:metric}). Designing the boundary conditions, we discussed the presence/absence of recurrences based on the reflection of correlation fronts. Additionally, our measurements reveal that the quench dynamics together with reflections from boundaries transform the initial thermal correlations into the so-called rainbow correlations at half of the recurrence time.

Our work opens the possibility for detailed studies of dynamics and correlations in an inhomogeneous metric. The ability to study the spatially resolved field $\phi$ together with the high level of control offered by the digital micromirror device, that shapes the spatio-temporal evolution of the averaged background density, offers a versatile platform. In particular, designing $v(z) \sim z^{-\kappa}$ would enable investigation of possible divergence of the signaling 
time for $\kappa<1$. 
This would shed light on the physics close to smoothed boundaries where corrections to Eq.~\eqref{eq:action} have to be taken into account.
Beyond simulating dynamics in curved spacetimes, the presented quantum field simulator can be used to study dynamics in inhomogeneous $1+1$-dimensional quantum fluids, which have attracted significant theoretical interest~\cite{Whitlock2003, MoraCastin2003, Cazalilla2004, Gritsev2007, Geiger2014, Dubail2017tonks, Dubail2017lightcones, BrunDubail2018igff, GLM2018, LangmannMoosavi2019, Moosavi2019, Ruggiero2019breathing, Murciano2019, Bastianello2020ill, RCDD2020, Dubail2017tonks, Dubail2017lightcones, LangmannMoosavi2019, Moosavi2019,Gluza2022,Moosavi2022} in general, and the inhomogeneous Luttinger-Liquid model with a spatially dependent $K(z)$, in particular, which is theoretically expected to exhibit a breaking of the Huygens-Fresnel principle~\cite{Gluza2020}. 
It is the hope that the present work stimulates such
further quantum simulations of curved geometries.

\emph{Acknowledgements}.---We would like to thank B.~Rauer, and T.~Schweigler for helpful discussions in the early stage of the project. This work has been supported by the DFG/FWF CRC 1225 'Isoquant',  the DFG/FWF Research Unit FOR 2724 ‘Thermal machines in the quantum world’, and the FQXi program on ‘Information as fuel’ 
ESQ Discovery Grant ‘Emergence of physical laws: from mathematical foundations to applications in many-body physics’ of the Austrian Academy of Sciences (O{\"A}W).
J.~E.~has also been supported by the DFG CRC 183 and the BMBF (MUNIQC-ATOMS), as well as by the EU's Horizon 2020 research and innovation program under grant agreement No.~817482 (PASQuanS). M.~G. acknowledges support through the start-up grant of the Nanyang Assistant Professorship of Nanyang Technological University, Singapore which was awarded to Nelly Ng. F.~C., F.~M., and J.~Sabino acknowledge support from the Austrian Science Fund (FWF) in the framework of the Doctoral School on Complex Quantum Systems (CoQuS). %
J.~Sabino acknowledges support by the Funda\c{c}{\~a}o para a Ci\^{e}ncia e Tecnologia, Portugal (PD/BD/128641/2017). 
G.~G. and S.~S. acknowledge support from the European Union’s Horizon 2020 research and innovation program under the Marie  Sk\l{}odowska-Curie grant agreement No.~101026667 and No.~101030988 respectively.

\emph{Author contribution}.---M.~T. and P.~S. performed the experiment with contributions by F.~C., S.-C.~J., J.~Sabino and F.~M.~M.~T.~and N.~S.~analysed the experimental data. S.~S., M.~G.~and N.~S.~provided the theoretical methodology and calculations with helpful suggestions from S.~E. and G.~G.. J.~Schmiedmayer and J.~E.~provided scientific guidance on experimental and theoretical questions. J.~Schmiedmayer conceived the experiment. All authors contributed to the interpretation of the data and to the writing of the manuscript.

\vspace{-0.5cm}

\bibliography{exp_LC_bibliography,ref,ref2,ref_SE}

\ 
 
\newpage
\onecolumngrid
\newpage

\section*{Supplemental Material}

Here we present additional information complementing the discussion in the main text. We subdivided it into two sections focusing on theoretical and experimental aspects, respectively.
Firstly, in Sec.~\ref{sec:Theory} we provide more details on the theoretical discussion of the front dynamics presented in the main text.
In Subsec.~\ref{sec:TLL}, we begin by reviewing the details about effective field theoretical modeling of tunneling-coupled ultra-cold gases.
In Subsec.~\ref{sec:shapesfronts}, we discuss correlation front dynamics in three qualitatively different situations: {\it i)} For an infinite space with a flat metric, {\it ii)} a flat metric but with hard walls and {\it iii)} a curved metric again with walls.
Secondly in Sec.~\ref{sec:expsetup}, we begin by describing the experimental setup in Subsec.~\ref{sec:setup}.
Subsequently in Subsec.~\ref{sec:correlations} we define the estimators of the second-moments of the velocity fields and finally, in Subsec.~\ref{sec:estimation} we describe our methodology for estimating the position of correlation fronts and extracting the front velocity.
These subsections link to figures presenting additional data analogous to those presented in the main text.


\section{Additional theoretical details
\label{sec:Theory}}

\subsection{Luttinger liquid description of tunneling-coupled atomic gases
\label{sec:TLL}}

As has been explained in several earlier works (see,
e.g., 
Refs.~\cite{Gritsev2007,Gritsev2007PRL,Schweigler_SG,Beck2018}),
the relative phase $\hat{\phi}(z)$ between two coupled parallel one-dimensional
gases of weakly interacting atoms at point $z$ is described at low energies by
the \emph{sine-Gordon Hamiltonian} 
\begin{equation}
    \hat{H}_{\mathrm{sG}}=\int\mathrm{d}z\biggl[\frac{\hbar^{2}\rho_0(z)}{4m}\left(\partial_{z}\hat{\phi}\right)^{2}+g\delta\hat{\rho}^{2}-2\hbar J\rho_0(z)\cos(\hat{\phi})\biggr]\label{eq:HsG}
\end{equation}
where $m$ is the atomic mass, $\rho_0(z)$ the expected atom density at position $z$, $g$ the
inter-atomic interaction, $J$ the tunnel coupling strength (controlled
by the tunneling of atoms between the two wells), and the $\delta\hat{\rho}(z)$
field is the relative density fluctuation about the mean density $n$ at $z$,
which is canonically conjugate to $\hat{\phi}(z)$, i.e., $[\hat{\phi}(z),\delta\hat{\rho}(z')]=-\mathrm{i}\delta(z-z')$. 
Note that throughout the Supplemental Material we adopted the simplified notation $g \equiv g_\mathrm{1D} $ (not to be confused with the metric determinant). 
{We assume that the mean density $\rho_0(z)$, which enters in the Hamiltonian as a space-dependent parameter, is only slowly varying with $z$ so that the hydrodynamic description remains  valid~\cite{Cazalilla}.} 

In the strong coupling limit and for sufficiently low temperature
equilibrium states, more specifically when the phase coherence length
\begin{align}
    \lambda_{T}(z) =\frac{2\hbar^{2}\rho_0(z)}{mk_{B}T}
\end{align}
is larger than the healing
length of the relative phase (typically a factor of four larger in the experiment, see Table~\ref{table:exppar})
\begin{align}
    \ell_{J}=\sqrt{\hbar/4mJ}\ ,
\end{align} 
the sine-Gordon model can be effectively approximated by the quadratic Klein-Gordon model. 
Introducing the sound velocity $v(z)$, \emph{Luttinger parameter} $K(z)$ and the Klein-Gordon quasi-particle mass, given in
terms of the microscopic parameters by
\begin{align}
    v(z) & =\sqrt{\frac{g\rho_0(z)}{m}}
    \label{appv}\ ,\\
    K(z) & =\frac{\hbar\pi}{2}\sqrt{\frac{\rho_0(z)}{mg}}
    \label{appK}\ , \\
    M_\text{KG} & =\sqrt{2\hbar mJ/g}\ ,
\end{align}
the Klein-Gordon Hamiltonian can be written as
\begin{equation}
    \hat{H}_{\mathrm{KG}}=\hat{H}_\text{TLL}+\frac{M_\text{KG}^2}{2}\int\mathrm{d}z\;v(z)^{2}\hat{\phi}^{2}\ , 
    \label{H_KG}
\end{equation}
where
\begin{equation}
    \hat{H}_{\mathrm{TLL}} 
    = \frac{\hbar}{2}
    \int\mathrm{d}z\,v(z)\biggl[\frac{K(z)}{\pi}\left(\partial_{z}\hat{\phi}\right)^{2}+\frac{\pi}{K(z)}\delta\hat{\rho}^{2}\biggr]
    \label{H_TLL}
\end{equation}
is the inhomogeneous \emph{Tomonaga-Luttinger liquid} (TLL) Hamiltonian~\cite{BrunDubail2018igff, Gluza2022, Moosavi2022}. For all of the above Hamiltonians, the equation of motion for the field $\hat{\phi}$ in the Heisenberg picture is
\begin{equation}
\partial_{t}\hat{\phi}(z,t)=(\mathrm{i}/\hbar)
[\hat{H}_{},\hat{\phi}(z,t)]=-\frac{2g}{\hbar}\delta\hat{\rho}(z,t)\ .
\label{eq:dtphi}
\end{equation}
This allows us to rewrite the TLL Hamiltonian as
\begin{align}\hat{H}_{\mathrm{TLL}}
   &=
\frac{\hbar}{2\pi} \int\mathrm{d}z\,K(z)\biggl[
v(z)\left(\partial_{z}\hat{\phi}\right)^{2}+v(z)^{-1}(\partial_t \hat \phi(z,t))^2\biggr]\ .
\end{align}
Finally, if we set the two non-trivial elements of the metric to be $g^{tt}=-v(z)^{-2}$ and $g^{zz}=1$ then we find $\sqrt{|\det (g_{\mu \nu})|}=v(z)$ which after identifying the canonical momentum leads to the Lagrangian density as stated in the main text. 
Based on the bosonisation formulas~\cite{Cazalilla} and restricting to the limit of weak inter-atomic
interaction, the particle density $\hat{\rho}$ and current $\hat{j}$
are 
\begin{align}
    \hat{\rho}(z) & =\rho_0(z)+\delta\hat{\rho}(z)\ ,\\
    \hat{j}(z) & =(\hbar \rho_0(z)/m) \partial_{z}\hat{\phi}(z)=: \rho_0(z)\hat{u}(z)
\end{align}
where we have defined the particle velocity field $\hat{u}$ as
\begin{equation}
    \hat{u}(z)=(\hbar/m)\partial_{z}\hat{\phi}(z)  \ .  
\end{equation}
Additionally, this equation of motion implies that both hydrodynamic fields $\hat{\rho}$ and $\hat{j}$
(equivalently $\delta\hat{\rho}$ and $\hat{u}$) can be expressed
in terms of time or space derivatives of the phase field $\hat{\phi}$. 
The post-quench dynamics is governed by $\hat{H}_{\mathrm{TLL}}$,
therefore the state remains Gaussian for all times.
The Heisenberg equations of motion of the fields $\hat{u}$ and $\delta\hat{\rho}$ are 
\begin{align}
\partial_{t}\hat{u}(z,t) & 
=-\frac{2g}{m}\partial_{z}\delta\hat{\rho}(z,t)\ ,  \\
\partial_{t}\delta\hat{\rho}(z,t) & 
=-\frac{1}{2}\partial_{z}(\rho_0(z)\hat{u}(z,t))\ .
\label{eq:EoM}
\end{align}

Combining the two equations we obtain the following equation for $\hat u$ 
\begin{align}
\partial_{t}^2 \hat{u}(z,t) & 
=\frac{g}{m}\partial_{z}^2 (\rho_0(z)\hat{u}(z,t))\ .
\label{eq:EoM_u}
\end{align}
Comparing the latter to the wave equation with inhomogeneous velocity $v(z)$, which is 
\begin{align}
\partial_{t}^2 \hat{u}(z,t) & = \partial_{z} [ {v(z)} \partial_{z} (v(z) \hat{u}(z,t))]
=\frac{g}{m}\partial_{z} \left[\sqrt{\rho_0(z)} \; \partial_{z} \left(\sqrt{\rho_0(z)}\hat{u}(z,t)\right)\right]\ .
\label{eq:inh_vel_u}
\end{align}
we observe that the two equations are different \cite{Gluza2022}. 
However, as we will explain later, the shapes of the light-cone fronts of correlations are the same in the two cases.

We now consider a quench from large to zero $J$ and we wish to characterize the time-evolved correlations of $\hat{u}(z)$, which can be done
following the arguments presented in Ref.~\cite{LongGaussification}. 
The initial state is a thermal state of $\hat{H}_{\mathrm{KG}}$.
This means that two important simplifications hold. First, the initial
state is Gaussian with zero mean, i.e., the only non-trivial correlations
of $\delta\hat{\rho}(z)$ and $\hat{u}(z)$ are the two-point correlations;
higher order correlations are given in terms of the latter by means
of Wick's theorem. Second, the initial state has a finite correlation length, $\xi\sim1/\sqrt{J}$. For sufficiently large $J$, the initial correlation length is short compared to the system size $L$, providing the length scale separation that is necessary for the observation of the light-cone fronts.

\subsection{Shapes of correlation fronts
\label{sec:shapesfronts}}
\subsubsection{Infinite space with flat metric}

We first focus on the case of flat metric in infinite space, which is realized if $\rho_0$ (consequently also the velocity $v$) is constant and the system size is infinite i.e. we are in the thermodynamic limit. 
The Heisenberg equations of motion (\ref{eq:EoM}) above can be combined  into two decoupled equations for $\hat{u}$
and $\delta\hat{\rho}$
\begin{align}
\partial_{t}^{2}\hat{u}(z,t) & =v^{2}\partial_{z}^{2}\hat{u}(z,t)\ , \label{equwave}\\
\partial_{t}^{2}\delta\hat{\rho}(z,t) & =v^{2}\partial_{z}^{2}\delta\hat{\rho}(z,t)\ ,
\end{align}
where $v=\sqrt{g\rho_0/m}$. 
Let us focus on the equation satisfied by $\hat u$ and solve the corresponding initial value problem in infinite space. 
The initial conditions are given by $\hat{u}(z,0)$ and the time derivative
\begin{align}
\left.\partial_{t}\hat{u}(z,t)\right|_{t=0} & =-\frac{\hbar\pi v}{mK}\partial_{z}\delta\hat{\rho}(z,0)=-\frac{2g}{m}\partial_{z}\delta\hat{\rho}(z,0).
\end{align}
The general solution $\hat{u}(z,t)$ for arbitrary initial conditions is 
given by
\begin{align}
\hat{u}(z,t) & =\frac{1}{2}\left(\hat{u}(z-vt,0)+\hat{u}(z+vt,0)+\frac{2g}{mv}(\delta\hat{\rho}(z-vt,0)-\delta\hat{\rho}(z+vt,0)\right)
\label{equzt}
\end{align}
which can be written in the shorthand form
\begin{align}
\hat{u}(z,t) & =\sum_{\sigma=\pm1}\hat{\chi}_{\sigma}(z+\sigma vt)\label{eq:solution0}
\end{align}
if we define the left- and right-moving chiral fields
\begin{align}
\hat{\chi}_{\sigma}(z) & :=\frac{1}{2}\left(\hat{u}(z,0)-\sigma\frac{2g}{mv}\delta\hat{\rho}(z,0)\right )\ .
\label{eq:chiral}
\end{align}
From the above we find that the time evolved correlations can be
expressed in terms of initial ones as
\begin{align}
\left\langle \hat{u}(z,t)\hat{u}(z',t)\right\rangle  =\sum_{\sigma_{1},\sigma_{2}=\pm1}\left\langle \hat{\chi}_{\sigma_{1}}(z+\sigma_{1}vt)\hat{\chi}_{\sigma_{2}}(z'+\sigma_{2}vt)\right\rangle \ .
\label{eq:solution_correlations}
\end{align}
\begin{figure*}[ht!]
    \centering
    \includegraphics[height=.45\linewidth]{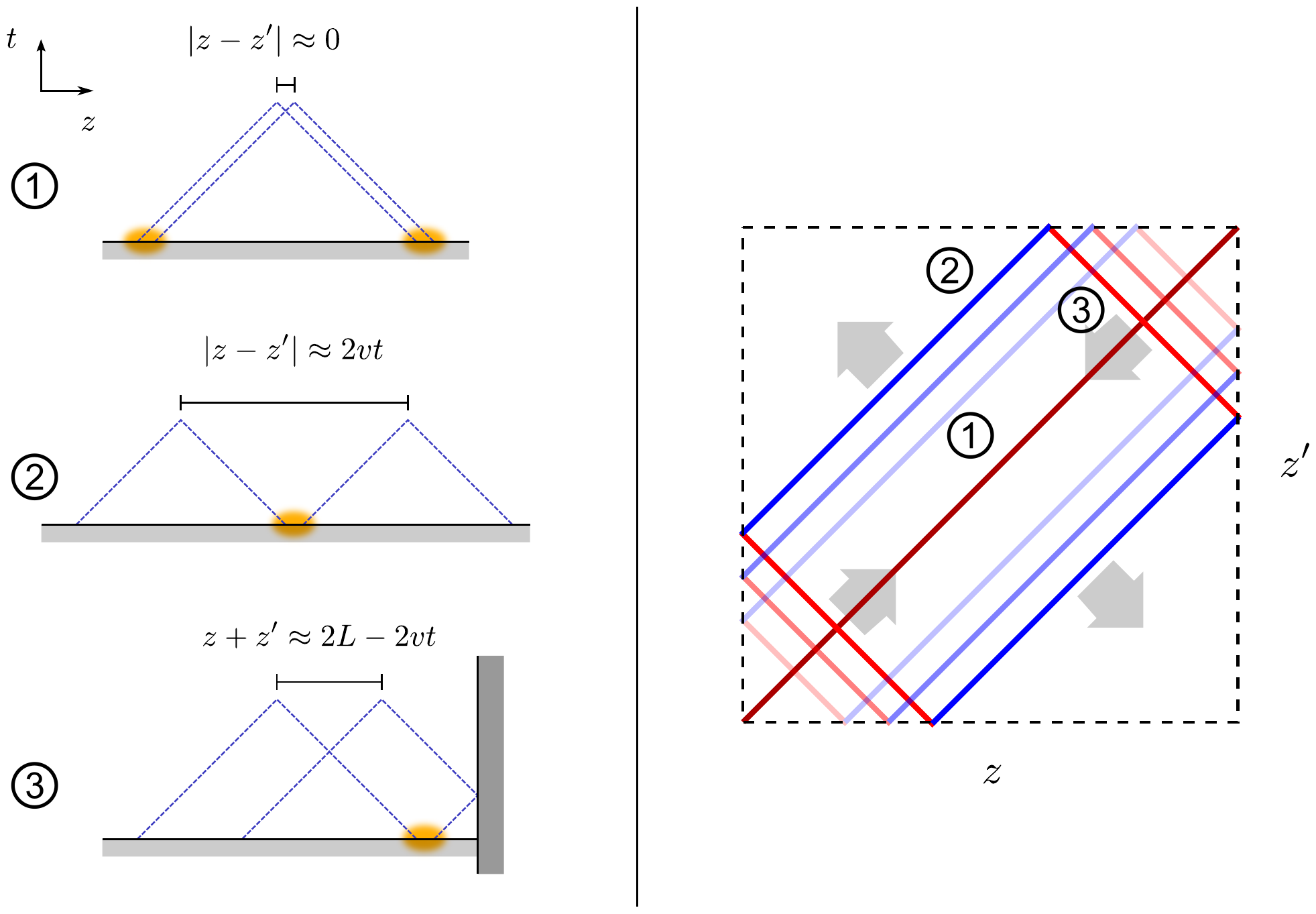}
    \caption{Illustrative explanation of the correlation propagation fronts in a finite box with a flat metric. 
    Equilibrium states of massive fields are characterised by correlation functions that decay exponentially with the distance, and here we consider initial states of this type with a correlation length $\xi$ much smaller than the system size $L$. 
    Following the quench to the massless phase, the time evolution of correlations between two points in space can be understood by backwards tracing the light-cone trajectories of left- and right-moving chiral fields. Correlations are significant only if the light-cone projections of the two points to the initial time surface are close to each other, i.e., at a distance of the order $\xi>0$. The apex of the cones signifies that there are two sources of measured correlation, one from the left and one from the right. The gray solid baseline signifies the initial state and the orange shape the extent of the initial correlation length. We can distinguish three cases: 
    {\bf (1)} Fields propagating to the same direction: these contribute only to diagonal correlations. 
   {\bf (2)} Fields propagating to opposite directions: these give rise to correlation fronts that appear parallel to the diagonal in the covariance matrix and move away from it at speed $2v$.
    {\bf (3)} One of the propagating fields reflected at one of the boundaries: this case results in fronts perpendicular to the diagonal that move `inwards'. 
    \label{fig:illustration1}}
\end{figure*}

It is straightforward to see that time evolved correlations between two points $z, z'$ at time $t$ can be
traced back to initial ones, since they originate from correlations between the past light-cone projection points $z\pm v t$ and $z'\pm v t$. 
Given that initial correlations are of short range,
there are only two possibilities for which the above sum can give
a non-negligible contribution: either $\left|z-z'\right|\approx0$
(in which case $\sigma_{1}=\sigma_{2}$), or $\left|z-z'\right|\approx2vt$
(in which case $\sigma_{1}=-\sigma_{2}$). 
The first case (depicted as case (1) in Fig.~\ref{fig:illustration1}) contributes only to correlations at nearby points for any time after the quench, while the second case (case (2) in Fig.~\ref{fig:illustration1}) results in the propagating correlation front which effectively travels at speed $2v$. 
By $\,\approx\,$ above, we more precisely
mean that the difference of the two sides is smaller than or of the same order as the initial correlation length $\xi\sim 1/M_\mathrm{KG}$ in absolute value.

{Note that in the experiment, for a quench from the Klein-Gordon to the Luttinger liquid with an initial correlation length that is much shorter than the system size $L$, the initial correlations of the density fluctuation field dominate over those of the velocity field~\cite{LongGaussification}. This can be seen from the ratio of their momentum mode variances, which is 
\begin{equation}
    \frac{\langle\delta\hat{\rho}_n^2\rangle}{\langle\hat{u}_n^2\rangle} \propto \frac{v^2k_n^2+M_\mathrm{KG}^2v^4/\hbar^2}{v^2k_n^2} \sim (L/\xi)^2 \gg 1
    \quad \text{for small } n \ .
\end{equation}
From (\ref{eq:solution_correlations}) and (\ref{eq:chiral}) we therefore see that the correlations after the quench are dominated by the contribution of initial correlations of $\delta\hat{\rho}$, which comes with a positive sign when $\left|z-z'\right|\approx0$ (since $\sigma_{1}=\sigma_{2}$) and with a negative sign when  $\left|z-z'\right|\approx2vt$ (since $\sigma_{1}=-\sigma_{2}$ in this case). This explains the negative sign of the propagating correlation fronts observed in the experiment.
}

\subsubsection{Finite box with flat metric and hard-wall boundaries}

Let us now consider the case of a finite system of size $L$ and assume that the confining potential is a hard-wall well. This means that
the particle current $\hat{j}(z)$ vanishes at the edges $z$ for all times
$t$, i.e., the boundary conditions imposed at $z=0,L$ are
of the Dirichlet type for the field $\hat{u}$ (i.e. Neumann type for the phase field) 
\begin{align}
    \left.\hat{u}(z,t)\right|_{z=0,L} & =0\ .\label{eq:BC}
\end{align}
The initial conditions of equation \eqref{equwave} and the domain of the solution $\hat{u}(z,t)$ are now restricted
to the spatial interval $[0,L]$. 
The boundary conditions can be easily
accommodated for by a simple modification of the previous problem
by extending the initial conditions beyond the interval
$[0,L]$ to infinite space: first, we reflect the initial data at the two boundaries and, second, we periodically repeat them with period $2L$. 
We can then use the infinite space solution above. 
Specifically, the solution can be still expressed in the form (\ref{eq:solution0}) as before, but to ensure that \eqref{eq:BC} is satisfied the chiral fields must now satisfy the equations
\begin{align}
	\hat{\chi}_{\sigma}(z) & = - \hat{\chi}_{-\sigma}(-z)\ , \label{eq:BC_chi1} \\
	\hat{\chi}_{\sigma}(z) & = - \hat{\chi}_{-\sigma}(2L-z) \ . \label{eq:BC_chi2}
\end{align}
which prescribe the extension of the initial conditions outlined above. 

In practice what this means is that, while tracing the light-cone projection points as before, we have to keep track of the reflections at the boundaries and the corresponding sign changes associated with the Dirichlet boundary conditions. 
The resulting solution can be written as
\begin{align}
    \hat{u}(z,t) = \sum_{\sigma=\pm1} \varsigma_{\sigma}(z,t)
    \hat \chi_{\sigma\varsigma_{\sigma}(z,t)}(\left|\zeta_{\sigma}(z,t)\right|)\ .
    \label{eq:EoM_box_solution}
\end{align}
where 
\begin{align}
    \zeta_{\sigma}(z,t) =  (z+\sigma v t)\operatorname{mod}_{(-L)} (2L)\ ,
\end{align}
and 
\begin{align}
    \varsigma_{\sigma}(z,t) = \operatorname{sign}\!\left(\zeta_{\sigma}(z,t)\right) \ ,
\end{align}
with $\; x \operatorname{mod}_{d} p \; $ being the $\operatorname{mod}$ function with real arguments and offset $d$  \begin{equation}
     x \operatorname{mod}_{d} p = ((x-d)\operatorname{mod} p)+d = 
x - p \left\lfloor \frac{x-d}{p} \right\rfloor \ .
\end{equation}
From  (\ref{eq:BC_chi1}) and (\ref{eq:BC_chi2}) it can be verified that the above solution satisfies the boundary conditions (\ref{eq:BC}) for all times. 

The above formula (\ref{eq:EoM_box_solution}) can be equivalently derived using the method of images, i.e., by first placing an opposite sign image at the reflection $-z$ of point $z$ with respect to the left boundary and, second, placing infinite copies of these images periodically in space with period $2L$. We then use the infinite space solution (\ref{eq:solution0}) with the infinite set of alternating sign images as sources and restrict it to the domain $[0,L]$.

\begin{figure*}[ht!]
    \centering
    \includegraphics[width=\linewidth]{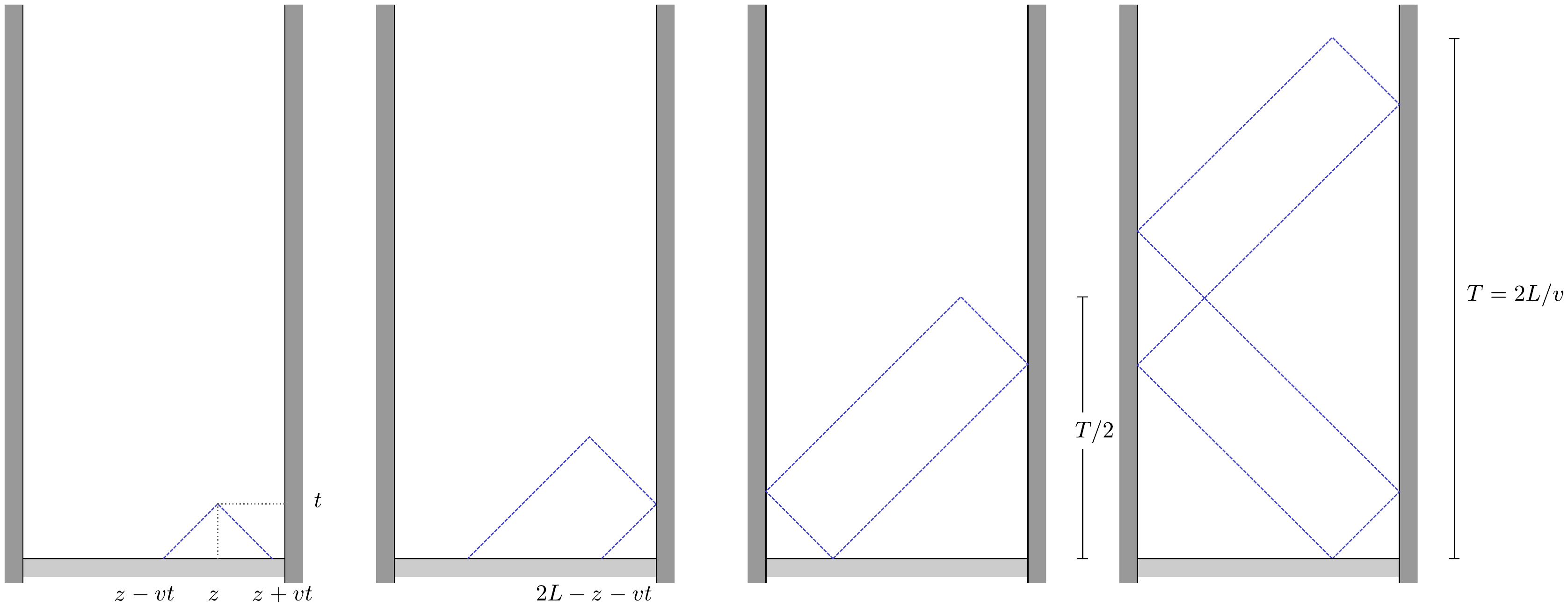}
    \caption{Geometric explanation of Eq.~(\ref{eq:EoM_box_solution}). The positions from which the two chiral components of the field $\hat{u}$ at position $z$ and time $t$ originate are $z\pm v t$, if $t$ is smaller than the first reflection time, $t<\text{min}(t_L,t_R)$ with $t_L = z/v, t_R = (L-z)/v)$ corresponding to the reflection times at the left and right boundaries, respectively. After reflection at the left or right boundary, the origins of the reflected chiral components are the points $-z+vt$ or $2L-z-vt$, respectively. The first recurrence of the initial state occurs at $t=T=2L/v$. However if the initial state is symmetric under space reflection $z\to L-z$ the correlations will exhibit a recurrence at $t=T/2$. \label{fig:lightcone_1}}
\end{figure*}
Note that at time $t=T$ where 
\begin{align}
    T & =\frac{2L}{v}\label{eq:Trec}
\end{align}
the solution is identically equal to the initial value $\hat{u}(z,T)=\hat{u}(z,0)$ for all points $z$ in $[0,L]$. 
The time $T$ is the period for a complete recurrence of the initial state. 
At $t=T/2$ the solution is equal to the initial value at the point corresponding to a reflection with respect to the middle point  $\hat{u}(z,T/2)=\hat{u}(L-z,0)$.

Considering the effect of the boundaries
on the calculation of the time evolved correlations between two points, \eqref{eq:EoM_box_solution} 
practically means that, in addition to the two possibilities $\left|z-z'\right|\approx 0$ or $2vt$ mentioned earlier for the infinite space case, which are now applicable only for sufficiently short times, there is one additional possibility giving
a non-negligible contribution: this is when one or both of the backwards propagating light-cone trajectories get reflected once or more times at the boundaries. For example, if only one of the trajectories gets reflected then the condition is $(z+z')\approx 0$ if the reflection is at the left boundary, and $(z+z')\approx 2L-2vt$ if the reflection is at the right boundary (case (3) in Fig.~\ref{fig:illustration1}). 
The three different cases are illustrated in Fig.~\ref{fig:illustration1}.

The above discussion shows that the time evolved correlation functions
are significant only close to the geometric locus of the points satisfying one of the three equations: $\left|z-z'\right|=0$, $\left|z-z'\right|=2vt$,
or $\left|z+\sigma z'\right|=|2nL+2\sigma vt|$ for integer $n$ and $\sigma=\pm1$. In the $(z,z')$ plane
and at a fixed time $t$, this locus is the union of the diagonal
line $z=z'$ and a rectangle with vertices along the lines
$z_{1,2}=0,L$ and edges parallel and perpendicular to the diagonal,
as illustrated in Fig. \ref{fig:illustration1}. 
The sign change of the correlations after the reflection at the boundaries is explained by the sign change of the propagating fields under the Dirichlet boundary conditions \eqref{eq:BC}. 

At $t=T$ the correlations are identical to the initial ones. However, because of the symmetry of both the post-quench Hamiltonian and the initial state under spatial reflection with respect to the middle, the period of the recurrence of correlations is $T/2$ \cite{Recurrence}. At $t=T/4$ the correlations are significant only along the diagonal and the anti-diagonal $z'=L-z$, which explains the characteristic `rainbow' configuration observed in the experiment. 
These are the main points needed to understand the patterns observed in the experimental data for the correlations.

\begin{figure*}[ht!]
    \centering
    \includegraphics[width=.8\linewidth]{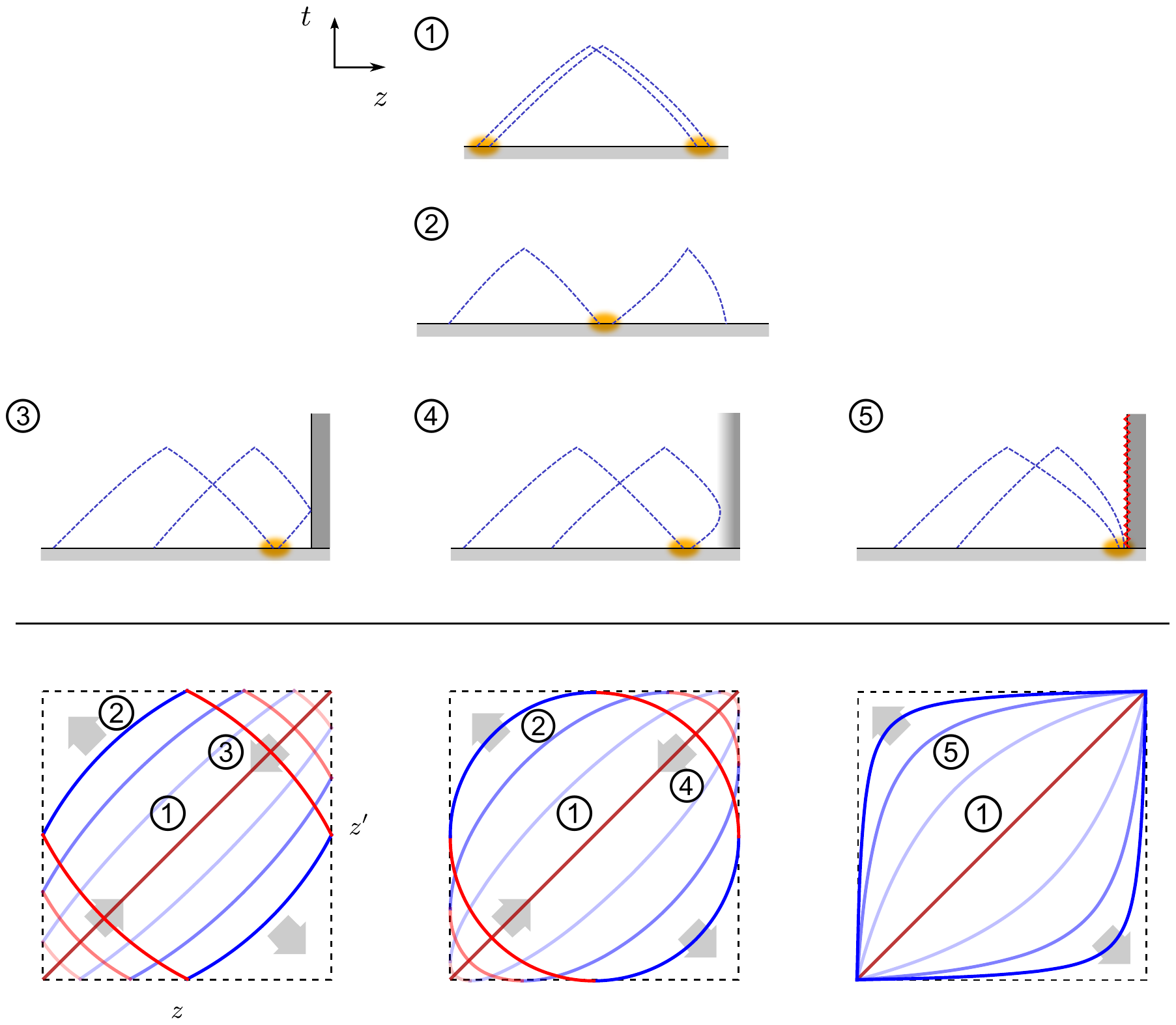}
    \caption{
    Similar to Fig.~\ref{fig:illustration1} but for a box with a curved metric. 
    As in the flat case, correlation dynamics can be derived by tracing the light-cone trajectories, which are now curved due to the spatial variation of the velocity $v(z)$, resulting in curved correlation fronts. Apart from the previously discussed cases {\bf (1-3)}, additional possibilities alternative to {\bf (3)} arise when $v(z)$ goes to zero at the edge {\bf (4-5)}. 
If $v(z)$ decays sufficiently fast, then trajectories reach the boundary at finite time and have a turning point there {\bf (4)}. If instead it decays sufficiently smoothly, then light-cone trajectories never reach the boundary, therefore they are not reflected nor turn back, instead they approach asymptotically the boundary {\bf (5)}. 
    \label{fig:illustration2}}
\end{figure*}

\subsubsection{Finite box with inhomogeneous metric}

We are now ready to study the inhomogeneous metric case. Assuming that $v(z)>0$ for all points $z$ in the closed interval $[0,L]$, the derivation of time evolved correlations is similar to the flat metric case with hard walls, with the difference that the time needed for a local field to propagate from point $z$ to $z'$ equals $|s(z',z)|$ with 
\begin{equation}
    \tau(z',z)=\int_{z}^{z'}\frac{\mathrm{d} l}{v(l)} \ , 
    \label{tau}
\end{equation}
c.f. (\ref{eq:s}). 
Therefore, the shape of the correlation fronts is determined by the following problem: for a given time $t$ find all points $z,z'$ (assuming $z\leq z'$ without loss of generality) such that there exist a point $z_{0}$ between them ($z\leq z_{0}\leq z'$) so that
\begin{equation}
    \tau(z',z_{0}) = -\tau(z,z_{0})=t \ .
\end{equation}
To find for each $z$ the corresponding $z'$ as a function of $z$ and $t$, we eliminate $z_{0}$ from the above equations by subtracting them, obtaining
\begin{equation}
    \tau(z',z) = 2t \ .
\end{equation}
Similarly, for $z'\leq z$ we get the opposite sign on either
side. Therefore, the two cases together correspond to 
\begin{equation}
    \tau(z',z)=\pm2t \ ,
    \label{eq:front_curve}
\end{equation}
i.e., the distance between the two points is such that a particle would
need time $2t$ to travel from one to the other in either direction. 
If we denote by $\tau_0(z)$ the travel time from, say, the centre of the system $z_0=L/2$ as a reference point, to any other point $z$ 
\begin{equation}
    \tau_0(z):=\tau(z,z_0) \ ,
\end{equation}
then the solutions $z'_{\pm}$ of  (\ref{eq:front_curve}) as functions of $z$ are 
\begin{equation}
    z'_{\pm}=\tau_0^{-1}\left(\tau_0(z)\pm 2t\right) \ .
    \label{eq:curved_front1}
\end{equation}
Note that $\tau_0$ is invertible, since its derivative is a positive function. 
If any of these values is beyond the edges, $z'_+>L$ or $z'_-<0$, then we subtract from $2t$
the time it takes to go from $z$ to the edge and turn back
\begin{align}
    z''_+ & 
    = \tau_0^{-1}\left(2\tau_0(L)-2t-\tau_0(z)\right), \label{eq:curved_front2a} \\
    z''_- & 
    = \tau_0^{-1}\left(2\tau_0(0)+2t-\tau_0(z)    \right)\ . \label{eq:curved_front2b}
\end{align}
Equations (\ref{eq:curved_front1},\ref{eq:curved_front2a},\ref{eq:curved_front2b}) determine the shapes of the main and the reflected correlation fronts when the velocity does not vanish at any point (cases (1-3) in Fig.~\ref{fig:illustration2}).

If $v(z)$ vanishes at one point $z$, as can happen at the edges of the system, then there are additional possibilities for the behavior of the time evolved correlations. 
If $v$ goes to zero at the edges sufficiently fast, 
then the time needed for a chiral field to travel from a point in the bulk to the edge can be finite, but instead of reflection its trajectory would have a turning point there. 
As a result the correlation fronts are tangent to the edge lines (case (4) in Fig.~\ref{fig:illustration2}) instead of having a cusp. 
On the other hand, if $v$ goes to zero sufficiently slowly, 
then the time needed to reach the edge could be infinite. In this case there would be neither reflection nor turn at the edge, and trajectories would instead converge to the edge asymptotically for large times (case (5) in Fig.~\ref{fig:illustration2}).   
Recurrences should also be absent given that their emergence relies on boundary reflection or turn. As a result, in the large time limit the correlations between all pairs of points in the bulk should approach steady state values that are determined by initial correlations in the neighborhood of the boundary. 
More specifically, if 
\begin{equation}
    \lim_{z\to z_b}v(z)=0 
    \qquad \text{with } v(z) \sim |z-z_b|^\alpha,
\end{equation}
where $z_b$ is the edge coordinate, 
then for $\alpha < 1$ we have that the traveling time to the edge $\lim_{z\to z_b} \tau_0(z)$ is finite, while for $\alpha \geq 1$ it is infinite.

Note that this last behavior seems to describe also the case of a parabolic trap with no hard walls, shown in the main text. 
In the Thomas-Fermi limit the density profile
is $n(z)\sim\mu-V(z)=\mu-\tfrac{1}{2}\omega^{2}z^{2}$ and diagonalising
the corresponding inhomogeneous Luttinger liquid Hamiltonian (which can be done expanding in Legendre polynomials instead of Fourier modes as the eigenfunctions \cite{Petrov2004,Citro2008,Gluza2022}) suggests
that a significant quasi-recurrence should occur at time $T=2\pi R_{TF}/v_0=\pi L/v_0$ ~\cite{Gluza2022} where $R_{TF}$ is the Thomas-Fermi radius of the gas and $v_0$ the speed of sound at the midpoint. 
However, this is not observed in the experiment, which is a clear sign of suppression of the reflections at the edges. 
In fact, based on the above formula for the
density profile one would expect that the density vanishes at the points $z=\pm R_{TF}$, which however is not true since this formula
is only valid in the bulk of the system.
In the vicinity of the edges
$z\approx\pm R_{TF}$ the Luttinger liquid description breaks down due to the vanishingly low density. The density profile actually turns out to decay in a smooth way as shown by the \emph{Gross-Pitaevski equation} 
\cite{Stringari_Bose_gas_edges}. 
There is no simple theoretical description of the dynamics of quasi-particles close to the edge 
and we cannot resolve experimentally the phase profiles in this region due to the very low density. Nevertheless, the absence of recurrences is an indirect evidence that there is no quasi-particle reflection or turn, so it is reasonable to assume that their trajectories converge asymptotically to the edges as when the effective velocity profile decays slowly.


\subsection{Difference between the curved metric model and the inhomogeneous Tomonaga-Luttinger liquid model}

The inhomogeneous Tomonaga Luttinger liquid Hamiltonian describing our experimental system given by (\ref{H_TLL}) is {not} equivalent to the free massless quantum field theory in inhomogeneous background metric as defined by the Hamiltonian (\ref{H_KG}) of the main text for $M=0$. As anticipated, this is due to the spatially dependent Luttinger liquid parameter $K(z)$ which cannot be absorbed in a redefinition of the fields or of the coordinates~\cite{Gluza2022,Brun2018}. Note that the spatial dependence of $K$ is controlled by the same atomic gas parameter $\rho_0(z)$ as the sound velocity $v(z)$, which is the key quantity of interest in our analysis; in fact they are both proportional to $\sqrt{\rho_0(z)}$ therefore if $\rho_0(z)$ varies sufficiently to have visible effects on $v$ it must also have equally visible effects on $K$. However, as far as quench correlations are concerned, while the spatial dependence of $K$ does play a role on correlations, it turns out that it has no effect on the shape of correlation fronts. Indeed, as shown in 
Ref.~\cite{Gluza2022} for the present case $K(z)\propto v(z)$, the spatial dependence of $K$ gives rise to dispersive effects that break the Huygens--Fresnel principle, i.e., the property that information travels strictly on light-like trajectories. As a result, information propagation is characterized by long tails behind the information fronts. However, the front shape itself is unaffected and fully determined by $v(z)$ only. The analysis of \cite{Gluza2022} applies to the Green's functions encoding the general solution of the initial value problem of the Heisenberg equations of motion and carries over to the study of quench correlations.

\begin{figure*}
    \centering
    \includegraphics[width=\linewidth]{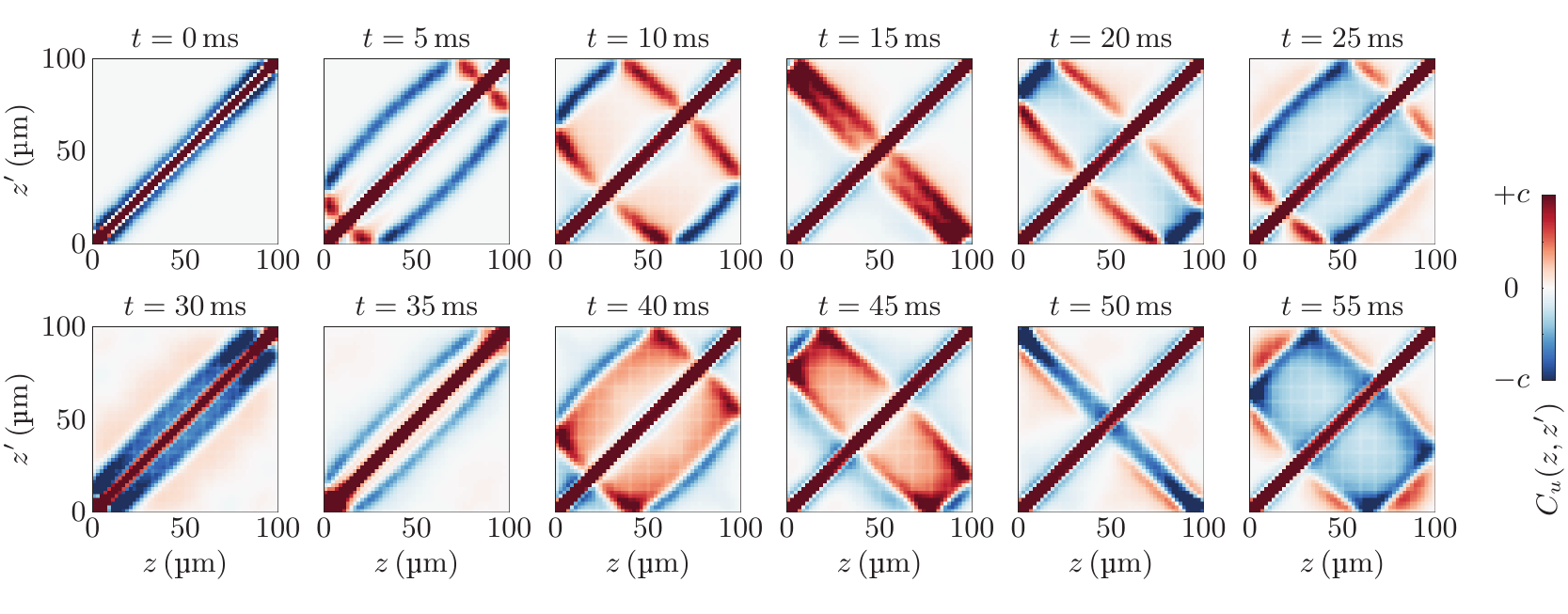}
    \caption{Example of how a KG to TLL quench can generate curved fronts. We consider a thermal KG state with temperature $T=\SI{50}{\nano\kelvin}$ and tunnel coupling $J=2\pi \times \SI{2.6}{\second}^{-1}$ computed with the Hamiltonian being parametrized by the inhomogeneous density profile with sharp edges as measured in the experiment at initial time $t=\SI{0}{\milli\second}$ (and smoothly refined by interpolation to match the discretization grid). The initial correlations have been propagated using the corresponding inhomogeneous TLL Hamiltonian.  Each plot presents the two point correlations of the velocity field, subjected to a Gaussian convolution accounting for the finite measurement resolution. The numerical simulation reproduces the appearance of the perpendicular correlation front and the outwards propagation of the anti-correlation front. }
    \label{fig:theoryfronts}
\end{figure*}

One way to explain the above observation is by comparing the Heisenberg equations of motion for the inhomogeneous TLL with $v(z),K(z)$ both proportional to the same function $\sqrt{\rho_0(z)}$ with those for the case where $v(z)$ is the same but $K$ is constant. In the former case we have 
\begin{align}
    \partial_{t}^{2} \hat{\phi}=v_0^{2} \partial_{z}\left(F(z) \partial_{z} \hat{\phi}\right) 
    \label{iTLL}
\end{align}
where we expressed both $v(z)$ and $K(z)$ in terms of the dimensionless function  $F(z):=\rho_0(z)/\rho_0(z_0)$ and $v_0=v(z_0)$ for some reference point $z_0$ \cite{Gluza2022}.
On the other hand, in the latter case we have
\begin{align}
    \partial_{t}^{2} \hat{\phi}=v(z) \partial_{z}\left(v(z) \partial_{z} \hat{\phi}\right) = v(z) v^{\prime}(z) \partial_{z} \hat{\phi}+v(z)^{2} \partial_{z}^{2} \hat{\phi}
    \label{inh_velocity_weq}
\end{align}
which is the inhomogeneous velocity wave equation. Eqs. \eqref{iTLL} and \eqref{inh_velocity_weq} are equivalent to \eqref{eq:EoM_u} and \eqref{eq:inh_vel_u} above, respectively, for the $\hat{u}$ field. 
Expressing $F(z)$ in the first equation (\ref{iTLL}) in terms of $v(z)$, we can bring it to a form comparable to (\ref{inh_velocity_weq}). Since $v(z)=v_0\sqrt{F(z)}$ we have
\begin{align}
    \partial_{t}^{2} \hat{\phi} = 2 v(z) v^{\prime}(z) \partial_{z} \hat{\phi}+v(z)^{2} \partial_{z}^{2} \hat{\phi}
    \label{iTLL_v2}
\end{align}
from which we notice that the difference between (\ref{iTLL_v2}) and (\ref{inh_velocity_weq}) is in the numerical coefficient of the first term. This shows that the two equations are indeed inequivalent, but at the same time shows that the highest derivative terms of these equations are the same. Given that this is what determines the motion of the fastest particles, we find that the first signal carrying the information of a local disturbance through the system travels equally fast in the two cases, therefore the same holds for the correlation fronts when starting from the same initial state.  
Fig.~\ref{fig:theoryfronts} shows an example of the formation of curved fronts in a KG to TLL quench described by (\ref{H_TLL}). In particular, note the bending over time of the anti-correlation front and the change of sign in the formation of the reflected front.

\section{Experimental setup, measurements and data analysis\label{sec:expsetup}} 
\subsection{Experimental setup and measurement protocol
\label{sec:setup}}

Our quantum field simulator consists of two parallel tunneling-coupled superfluids trapped by a combination of magnetic and optical dipole traps on an atom chip~\cite{reichel2011atom}. 
We cool down around $20000$ atoms of $^{87}$Rb in a highly anisotropic parabolic magnetic potential produced by the atom chip~\cite{Rauer2016}. The typical trapping frequency in the (tight) transverse and (shallow) longitudinal axis are $\omega_\perp / 2\pi = 1.4\, \si{\kilo \Hz}$ and $\omega_z / 2\pi = 7\, \si{\Hz}$ respectively. Using radiofrequency fields, a double-well potential can be realized in one of the transverse directions~\cite{Schumm05,Schweigler_SG}.
The single particle tunneling rate between the two wells, $J$, can be tuned by changing the amplitude of the radiofrequency field. Typical values of $J/2\pi$ in the strong coupling regime are around $1\, \si{\Hz}$. Adding an optical dipole potential in the longitudinal direction allows for obtaining arbitrary potentials in $z$ direction. To adjust this potential, we use a digital micro-mirror device (DMD) to shape a laser beam $\lambda =660\, \si{\nm}$, blue-detuned with respect to D$_2$ transition line of $^{87}$Rb~\cite{DMD}.

\begin{figure}[ht]
    \centering
    \includegraphics[width=\linewidth]{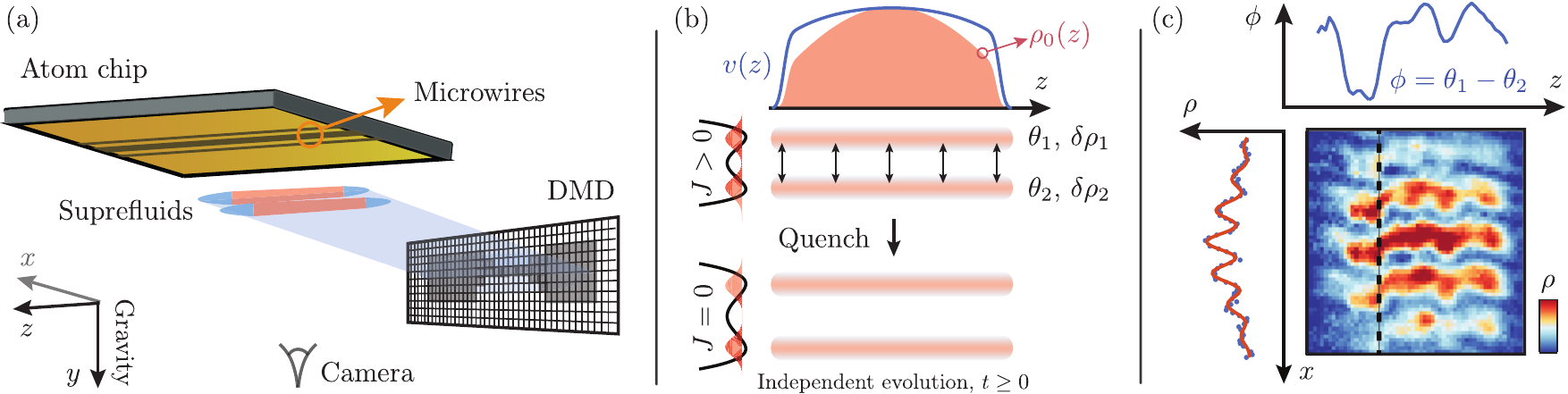}
    \caption{Schematics of the experimental setup, quench protocol, and the phase measurement. (a) Two tunneling-coupled superfluids are shown below an atom chip. The projected DMD pattern modifies the potential in the $z$ direction. (b) Initially, the gas is cooled down in a coupled DW potential. Each condensate ($j=1,2$) is described by $\psi_j(z) = \exp{(\mathrm{i}\theta_j(z))}\sqrt{\rho_0(z) + \delta \rho_j(z)}$. The tunnelling-coupling is then quenched to zero and initiates the independent evolution of the atoms in separated traps. The local speed of sound, $v(z)$, depends on the local density $\rho_0(z)$ which can modified. (c) After free fall and expansion, the two-dimensional projected density profile, $\rho(x,z)$ is obtained by absorption imaging. The spatially resolved relative phase $\phi(z)$ is determined from interference images. For every slice in $z$ direction, a cosine function multiplied by a Gaussian bell is fitted to extract the relative phase. The red curve in the figure is shown as an example at $z=z_0$.} 
    \label{fig:expsetup}
\end{figure}

Initially, the two cigar-shaped clouds are prepared in a strongly tunneling-coupled double well in the transverse direction. For the initial state, the dimensionless ratio between the thermal coherence length and the healing length of the phase, $q = \lambda_T / \ell_{J}$ is estimated to be about $5$ (see Table~\ref{table:exppar}). This ratio is directly related to $\left\langle \cos{\phi(z)} \right\rangle$ and can be estimated using simulations in classical field approximation~\cite{Thomas_thesis}. To perform the quench from coupled to uncoupled double wells, the tunneling rate is changed to zero by increasing the barrier height between the two clouds in $2\, \si{\ms}$ (see Fig.~\ref{fig:expsetup}(b)). After quenching, we let the two independent condensates evolve for different trapping times, $t$, and then we measure the relative phase between the two condensates, $\phi(z)$, by matter-wave interferometry. 

To probe, we release the atoms by switching off all the traps and take an absorption image from below after they freely fall for $15.6\, \si{\ms}$ and interfere with each other (see Fig. \ref{fig:expsetup}(c))~\cite{van2018projective}. Due to the destructive nature of absorption imaging, the measurement is repeated several times for every $t$.
The camera pixel size in the plane of atoms is $2\, \si{\micro\meter}$. modeling the imaging system with a Gaussian \emph{point spread function} (PSF), the estimated standard deviation of this function will be $\sigma_\mathrm{PSf} = 3\, \si{\micro\meter}$~\cite{Thomas_thesis}. 
To measure the temperature, we use the `density ripples' after long free expansions. The initial thermal phase fluctuations transform into density fluctuations after $11.2\, \si{\milli \s}$ of free expansion, allowing for extraction of temperature from their correlations~\cite{Manz2010DRthermometry, Moller2021NNthermometry}.

\subsection{Construction of the correlation function}
\label{sec:correlations}

\begin{figure}
    \centering
    \includegraphics[width=\textwidth]{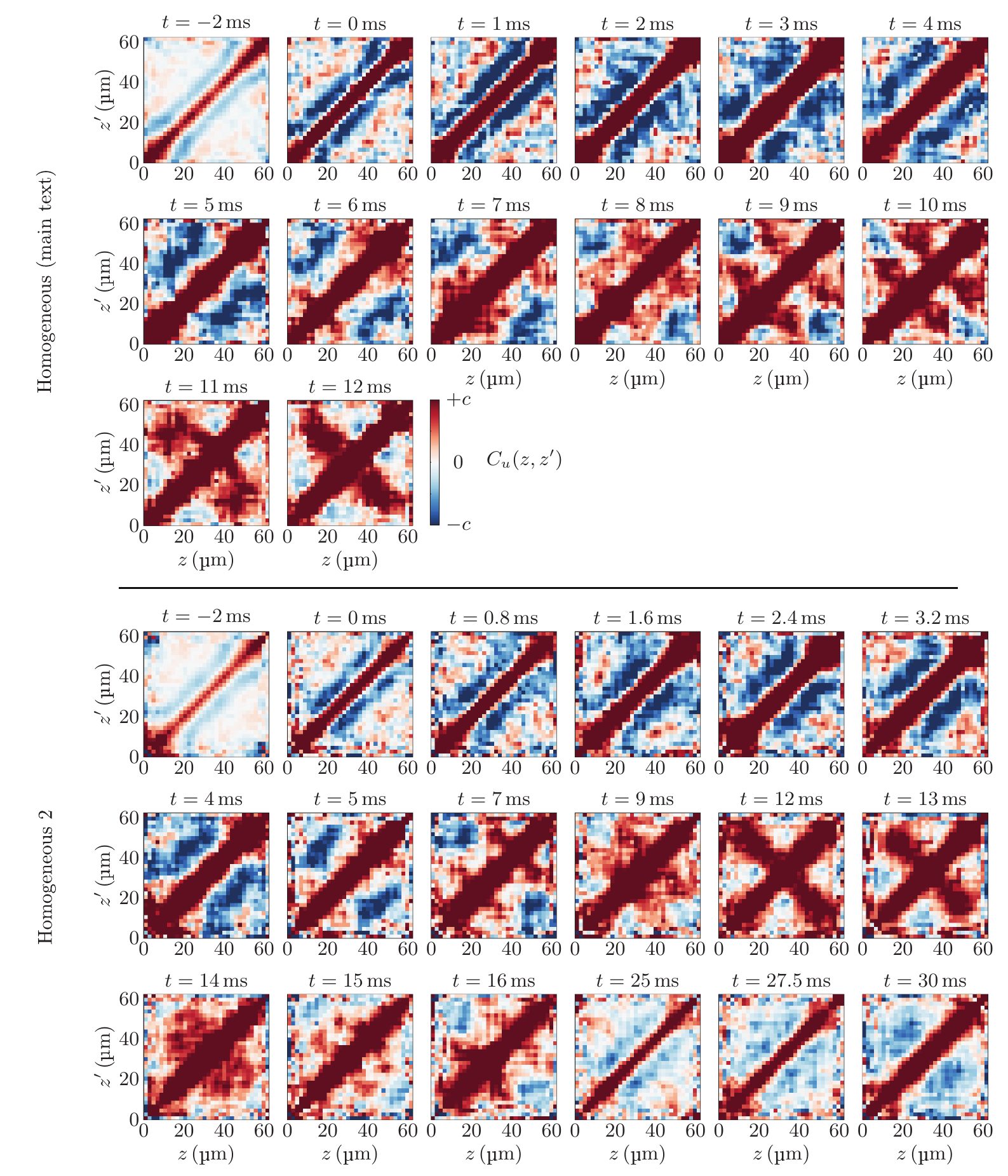}
    \caption{Measurement results of $C_u(z,z^\prime)$ at different times $t$ for two settings with homogeneous background density.}
    \label{fig:all_corr_flat}
\end{figure}

\begin{figure}
    \centering
    \includegraphics[width=\textwidth]{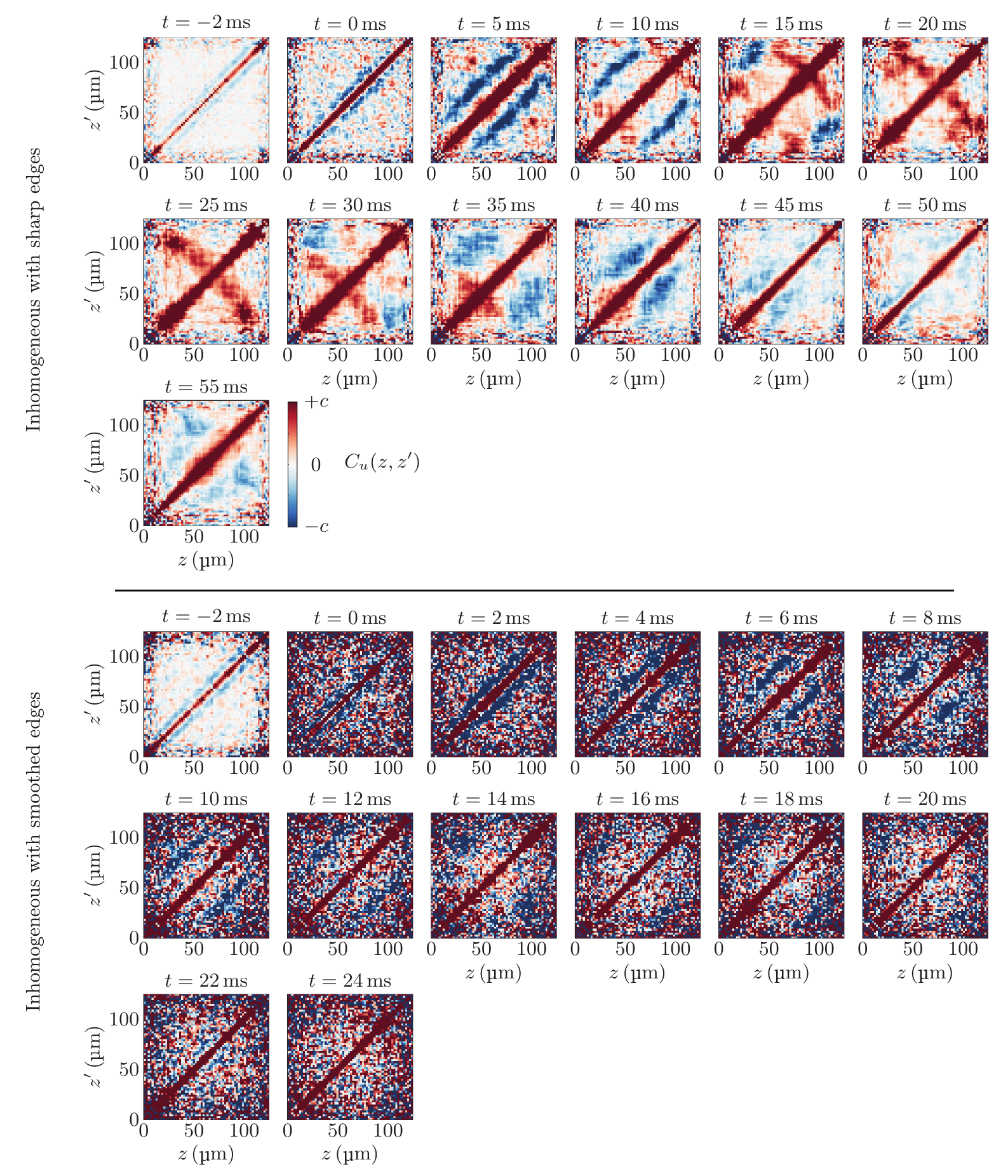}
    \caption{Measurement results of $C_u(z,z^\prime)$ at different times $t$ for two settings with inhomogeneous background density.}
    \label{fig:all_corr_curved}
\end{figure}

The velocity field, $\hat{u}(z)$ is estimated based on the finite difference of the relative phase:
\begin{align}
    \hat{u}(z)=(\hbar/m)\partial_z \hat{\phi}(z) \approx (\hbar/m)\Delta_{z}\hat{\phi}(z)\, ,
    \label{eq:velocityfielddisc}
\end{align}
where 
\begin{align}
    \partial_z \hat \phi(z) \approx \Delta_z \hat{\phi}(z) = \frac{\hat \phi(z+\Delta z) - \hat \phi(z)}{\Delta z} \, ,
    \label{eq:velocityfielddisc2}
\end{align}
with $\Delta z$ being the camera pixel size. Two-point correlation functions of the velocity field, $C_u(z,z^\prime) = \langle \hat{u}(z) \hat{u}(z^\prime) \rangle$, is calculated by taking the ensemble average over hundreds of experimental realizations. The measured correlation functions are shown in Figs.~\ref{fig:all_corr_flat}~and~\ref{fig:all_corr_curved} for flat and curved backgrounds respectively.

\subsection{Estimation of the front position and the average front velocity}
\label{sec:estimation}

\begin{figure}
    \centering
    \includegraphics[width=\textwidth]{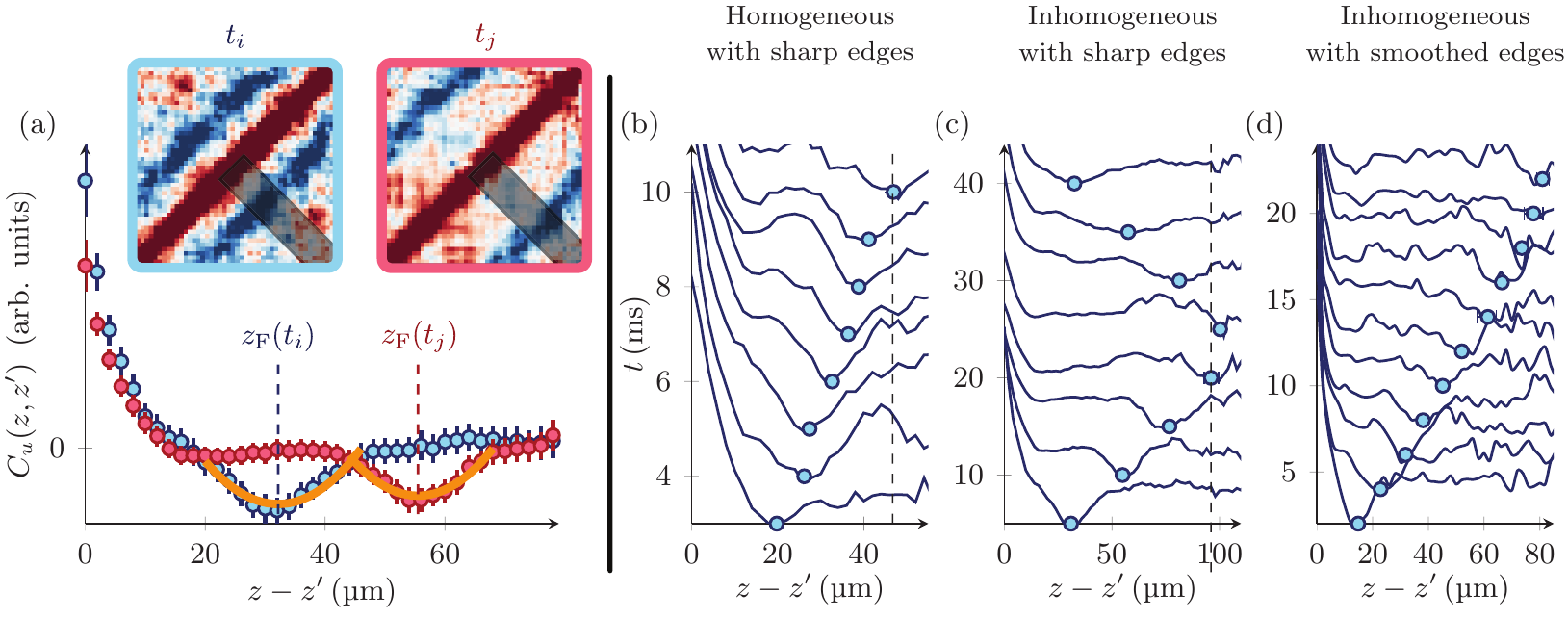}
   \caption{Estimation of the front position at different times. (a) for each time, we average along the diagonal to reduce the noise. The results are plotted in blue and red. The errorbars are the standard error of the mean. Orange curves are the parabolic fits. (b,c,d) The position of the fronts (light blue bullets) at different times for the three experimental settings presented in the main text. Dark blue curves are the anti-diagonal correlations similar to (a) (presented as lines only for illustration purpose).}
   \label{fig:front_pos}
\end{figure}

We evaluate a slice of $C_u(z,z^\prime)$ along the anti-diagonal, i.e., $z^\prime=-z$. We average over a few neighboring pixels to reduce the noise. The correlation front appears as a local minimum in this profiles as shown in Fig.~\ref{fig:front_pos}. In order to quantitatively estimate the position of the front, we fit a parabola around the minimum (orange curves in Fig.~\ref{fig:front_pos}). The minimum of the parabola is the estimated position of the front, $z_\mathrm{F}(t_i)$, at that particular time step, $t_i$. Bootstrapping with 2222 resampling is used to estimate $68\%$ confidence interval. We then calculate the average velocity at time $t$:
\begin{align}
    v_\mathrm{F}(t) = \frac{z_\mathrm{F}(t_i) - z_\mathrm{F}(t_j)}{t_i - t_j} \; ,\; t = \frac{t_i+t_j}{2}\, .
    \label{eq:velocityfielddisc3}
\end{align}

The measured velocities along with the theory calculations are shown in Fig~\ref{fig:velocity}. In order to quantify the agreement between the measured velocities and the theory calculations, we compute the reduced $\chi^2$ defined as
\begin{align}
    \chi^2_\mathrm{red} = \frac{1}{N-1} \sum_{i=1}^N \frac{(v_\mathrm{F}(t_i) - v_\mathrm{F}^\mathrm{theo}(t_i))^2}{\sigma_i^2} \, ,
    \label{eq:chi2}
\end{align}
where $N$ is the number of data points and $\sigma_i$ is the standard error of the mean. Here we use the half of the $68\%$ confidence interval obtained via bootstrapping. As a comparison, we compare the calculated $\chi^2_\mathrm{red}$ for the theory calculations with the reduced $\chi^2_\mathrm{red}$ obtained by considering a constant velocity (see Fig.~\ref{fig:velocity}).

\begin{table}[b]
\begin{center}
\begin{tabular}{ c c c c c c c}
\hline
 Background density  & $K_\mathrm{max}$ & $T\, (\si{nK})$ & $\lambda_T \, (\si{\micro\meter})$ & $\ell_{J}\, (\si{\micro\meter})$ & $q = \lambda_T / \ell_{J} $ \\ 
 \hline
 Homogeneous (main text)  & 47  & 51 & 19 & 4.0 & 4.7 \\  
 Inhomogeneous with sharp edges  & 68  & 86 & 16 & 3.1 & 5.2\\
 Inhomogeneous with smoothed edges  & 44 & 41 & 19 & 2.5 & 7.6 \\  
 Homogeneous 2 & 50 & 48 & 18 & 3.2 & 5.6 \\  
 \hline
\end{tabular} 
\end{center}

\caption{For every setting relevant parameters are listed: $K_\mathrm{max}$ is the Luttinger parameter calculated based on the maximum density, $T$ the temperature, $\lambda_T$ the thermal coherence length and $\ell_{J}$ the healing length of the phase. \label{table:exppar}}
\end{table}

\begin{figure}
    \centering
    \includegraphics[width=\textwidth]{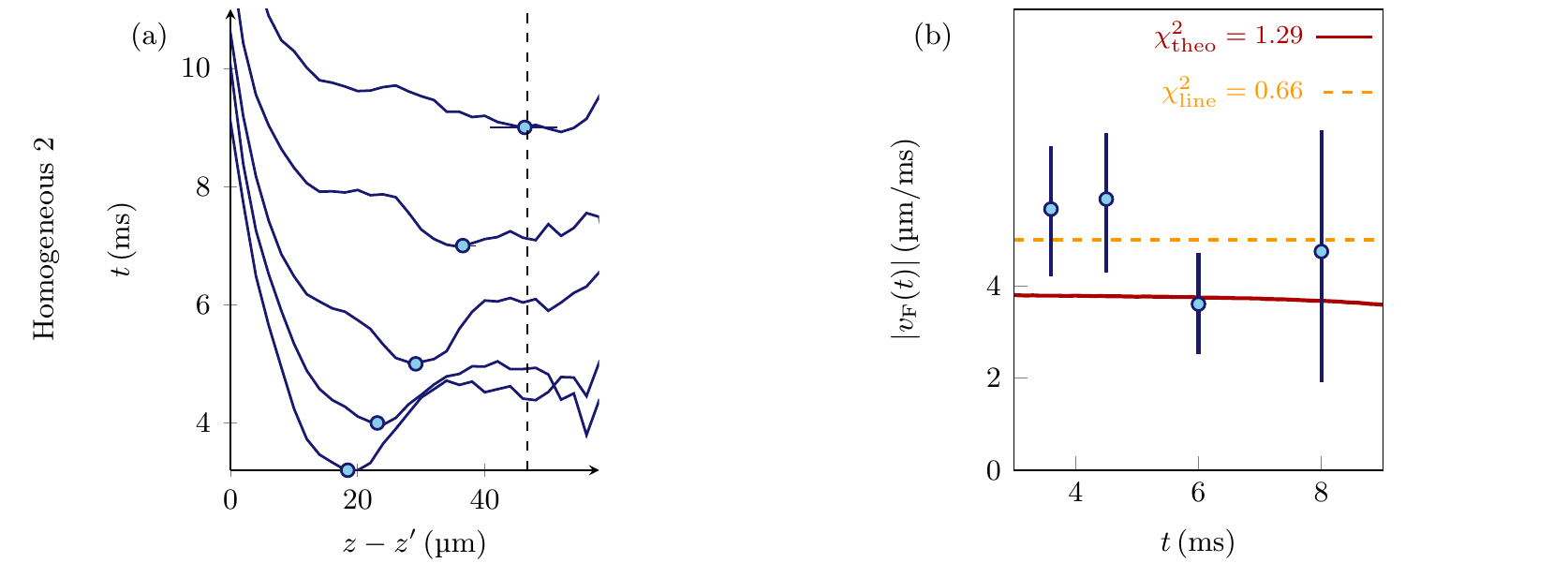}
   \caption{Data for the second measurement with a homogeneous density presented as "Homogeneous 2" in Fig.~\ref{fig:all_corr_flat} and Table~\ref{table:exppar}). (a) Estimation of the front position at different times, and (b) the average front velocity. See Figs.~\ref{fig:front_pos} and~\ref{fig:velocity} for details.}
   \label{fig:extendeddata}
\end{figure}


\end{document}